![polymers logo]

*Article*

# Impact of acid hydrolysis on morphology, rheology, mechanical properties, and processing of thermoplastic starch

**Saffana Kouka [1], Veronika Gajdosova [1], Beata Strachota [1], Ivana Sloufova [2], Radomir Kuzel [3], Zdenek Stary [1], Miroslav Slouf [1,*]**

[1]   Institute of Macromolecular Chemistry of the Czech Academy of Sciences, Heyrovsky Sq. 2, 16206 Prague, Czech Republic; kouka@imc.cas.cz; gajdosova@imc.cas.cz; beata@imc.cas.cz; stary@imc.cas.cz; slouf@imc.cas.cz

[2]   Charles University, Faculty of Science, Department of Physical and Macromolecular Chemistry, Hlavova 2030, 128 40 Prague 2, Czech Republic 2; ivana.sloufova@natur.cuni.cz

[3]   Charles University, Faculty of Mathematics and Physics, Ke Karlovu 5, Praha 2 121 16, Czech Republic; radomir.kuzel@matfyz.cuni.cz

*   Correspondence: Miroslav Slouf; slouf@imc.cas.cz

**Abstract:** We modified native wheat starch using 15, 30, and 60 min of acid hydrolysis (AH). The non-modified and AH-modified starches were converted to highly-homogeneous thermoplastic starches (TPS) using our two-step preparation protocol consisting of solution casting and melt-mixing. Our main objective was to verify if the AH can decrease the processing temperature of TPS. All samples were characterized in detail by microscopic, spectroscopic, diffraction, thermomechanical, rheological, and micromechanical methods, including *in situ* measurements of torque and temperature during the final melt-mixing step. The experimental results showed that: (i) the AH decreased the average molecular weight preferentially in the amorphous regions, (ii) the lower-viscosity matrix in the AH-treated starches resulted in slightly higher crystallinity, and (iii) all AH-modified TPS with less viscous amorphous phase and higher content of crystalline phase exhibited similar properties. The effect of the higher crystallinity predominated at laboratory temperature and low deformations, resulting in slightly stiffer material. The effect of the lower-viscosity dominated during the melt mixing, where the shorter molecules acted as a lubricant and decreased the *in situ* measured processing temperature. The AH-induced decrease in the processing temperature could be beneficial for energy savings and/or possible temperature-sensitive admixtures to TPS systems.

**Keywords:** thermoplastic starch; low viscosity; melt mixing; processing temperature

## 1. Introduction

Starch is the primary carbohydrate reserve in plants, composed of two polysaccharides: amylose and amylopectin [1]. Amylose is a linear polymer linked by α-1,4 glycosidic bonds, while amylopectin is formed by highly branched molecules with α-1,4 linkages in the backbone and α-1,6 linkages at the branching points [2]. Molecular weights of both polymers are very high: ca $10^6$ g/mol for amylose and ca $10^8$ g/mol for amylopectine [3–6]. In the native starch granules, amylose is localized in the amorphous regions, whereas amylopectin is in semicrystalline regions [7,8]. Due to its high molecular weight, the native starch cannot be processed by melt-mixing like other common polymers, because it decomposes before melting [9]. To overcome this limitation, starch is typically mixed with low molecular weight plasticizers such as water, glycerol, citric acid, and/or urea [10,11]. In our previous work, we demonstrated that oligomers such as maltodextrin can be added as lubricants that decrease starch processing temperature [2].

When starch is heated in the presence of plasticizers and subjected to shear forces, it undergoes gelatinization — a process in which the starch granules are partially disrupted and merged [8,12]. This transform the native starch to thermoplastic starch (TPS) that can be processed similarly to conventional thermoplastic polymers. TPS can be prepared by solution casting in the excess of water (SC; [13]), by melt-mixing (MM; [14]), or by the combination of both protocols (SC+MM; [15]). TPS must be prepared at elevated temperatures, as the starch gelatinization generally occurs above 100 °C, regardless of the water content [16]. However, even a moderate reduction of the processing temperature would be advantageous for the energy-saving reasons. One of possible strategies to decrease the processing temperature is the lowering of starch molecular weight, $M$. The polymers with lower $M$ exhibit lower viscosity and require lower temperatures to flow [17].

The reduction of the starch molecular weight can be achieved through physical methods, such as hydrothermal treatment [18]) or ultrasonication [19,20]), by means of chemical processes (typically acid hydrolysis [21]), or with enzymatic methods [22]. The acid hydrolysis (AH) is widely employed in the food industry to tailor starch's physicochemical properties [23,24]. It is often conducted below the gelatinization temperature to preserve the granular structure of starch [25]. The AH proceeds in two phases: an initial rapid hydrolysis targeting the amorphous regions, followed by a slower phase where crystalline regions are hydrolyzed [26]. Consequently, acid-hydrolyzed starch often exhibits higher crystallinity than native starch [23,25]. The acid hydrolysis of starch also affects the rheological properties of the TPS: mild acid hydrolysis can enhance gel strength and stiffness, while prolonged hydrolysis reduces molecular weight, producing weaker gels [27]. Although the acid hydrolysis of starch is a well-established industrial process, particularly in the context of food applications [23,28], limited information is available on its role in plasticization and the resulting processing properties of thermoplastic starch prepared by melt mixing [29,30].

In this contribution, we subjected native wheat starch to acid hydrolysis (AH) for 0, 15, 30 and 60 min. The AH-starch was thermoplasticized using a single-step solution casting (SC), and the two-step protocol combining solution casting with melt mixing (SC+MM) which yields the highly homogeneous thermoplastic starch [15]. The prepared materials were characterized thoroughly by numerous microscopic, spectroscopic, diffraction, rheological, thermomechanical, and micromechanical measurements. The first question we asked was how the AH treatment influences morphology, homogeneity and mechanical performance of thermoplastic starch. The second question was if the acid hydrolysis can decrease the starch viscosity and, as a result, its processing temperature during melt mixing. The highly homogeneous starch with lower processing temperature would be a promising material for both technical applications in packaging and agriculture (energy savings during the processing) and medical applications in pharmaceutics (the high homogeneity to secure reproducibility and lower processing temperature to improve the stability of sensitive admixtures, such as antibiotics [31]).

## 2. Materials and Methods

### 2.1 Materials

The wheat starch powder used in this study (starch type A, amylose content ca 25 %) was supplied by Škrobárny Pelhřimov a.s. (Pelhřimov, Czech Republic). Anhydrous glycerol ($C_3H_8O_3$; >99%), hydrochloric acid, (HCl; 35%), and sodium bromide (NaBr; >99%) reagents were bought from Lach-Ner s.r.o. (Neratovice, Czech Republic).

## 2.2 *Preparation of hydrolyzed starch*

Acid hydrolysis was performed following a previously described method by [32] with a few minor modifications. A 40% starch slurry was prepared by dispersing starch in an aqueous 1M hydrochloric acid solution. The reaction was conducted under mechanical stirring and mild heating in a water bath maintained at 45°C. Hydrolysis durations were varied (15 minutes, 30 minutes, and 60 minutes) to evaluate the effects of reaction time. To terminate the reaction, the mixture's pH was adjusted to 7 using a 1M sodium hydroxide solution. The neutralized slurry was then washed thoroughly with distilled water, followed by filtration. The resulting material was dried in an oven at 45°C for 24 hours and subsequently cooled to room temperature and milled into a fine powder.

## 2.3 *Preparation of thermoplastic starch*

The thermoplastic starch (TPS) was prepared from the native starch powders (S) with various acid hydrolysis (AH) times. TPS was prepared by both single-step solution casting protocol (SC; section 2.3.1) and two-step protocol comprising solution casting followed by melt mixing (SC+MM; section 2.3.2). This approach is based on our previous work [15], with slight modifications described below. The TPS samples prepared in this study are summarized in Table 1.

**Table 1.** List of prepared samples.

| Native starches * | Thermoplastic starches ** | AH time (min) *** |
|---|---|---|
| S-AH-00min | TPS-AH-00min | 0 |
| S-AH-15min | TPS-AH-15min | 15 |
| S-AH-30min | TPS-AH-30min | 30 |
| S-AH-60min | TPS-AH-60min | 60 |

\* Native starches were all based on wheat starch type A, differing only in AH time.

\*\* Thermoplastic starches were prepared by both single-step SC and two step SC+MM protocol.

\*\*\* Acid hydrolysis (AH) was applied to original native starches, from which TPS were prepared.

## 2.3.1 TPS prepared by single-step solution casting

The starch powder (70 wt.%) were premixed with glycerol (30 wt.%) and distilled water (6 parts of water per 1 part of starch) with a magnetic stirrer in a beaker for 30 min at room temperature. The pre-mixed suspension was transferred to a mechanical stirrer, where it was heated to initiate the starch gelatinization. A significant increase in viscosity was observed at temperatures between 63°C and 70°C. The viscosity increase indicated the onset of gelatinization, and the mixture was stirred continuously for about 15 minutes until a homogeneous pudding-like consistency was obtained. Then the solution was cast onto a polyethylene (PE) foil to form a film with a thickness of approximately 2 mm. The thin film was left to dry at room temperature for three days to allow for the evaporation of residual water.

## 2.3.2 TPS prepared by two-step protocol: solution casting followed by melt mixing

The solution-casted and dried TPS films from the previous steps were processed by melt mixing using a twin-screw laboratory kneader (Brabender Plasti-Corder, Duisburg, Germany) to further increase their homogeneity [15,33]. The samples were mixed in the chamber preheated to 120 °C, using rotation speed 60 rpm for at least 8 min while recording the real processing temperature and torque moments. Subsequently the sample was compression molded into plaques with a thickness of 2 mm. This was achieved using a laboratory hot press (Fontijne Grotnes; Vlaardingen, Netherlands) in the following multi-step process: Initially, the material was pressed at 130 °C under a pressure of 50 kN for 2 min to deaerate. This was followed by pressing at the same temperature under 100 kN for 2 min. Finally, the press with the molded plaques was cooled with water while maintaining a pressure of 100 kN for approximately 10 min, until the room temperature was reached.

### 2.3.3 Storing of the samples at defined conditions

The final TPS plaques were stored in defined conditions: at room temperature in a desiccator over a supersaturated solution of sodium bromide, which yields relative humidity = 57 %. The samples were in the desiccator all time, being removed just before the measurements of their properties by the characterization methods described below. Our experience showed that storing the samples at well-defined humidity leads to more reproducible results of mechanical and rheological measurements, even if the storage at ambient conditions is possible as well.

### 2.4 Characterization methods

#### 2.4.1 Light and electron microscopy

Morphology and homogeneity of TPS samples was checked by light microscopy (LM), polarized light microscopy (PLM), and scanning electron microscopy (SEM). The thin sections (thickness 5 μm) for LM and PLM were prepared with a rotary microtome RM 2255 (RM 2255; Leica, Vienna, Austria). The sections were placed in a thin oil layer between the microscopic glasses and observed in transmitted light (LM) or polarized transmitted light (PLM) in a microscope Nikon Eclipse 80i (supplied by Laboratory Imaging, Praha, Czech Republic). The fracture surfaces for SEM observations were prepared in liquid nitrogen. The specimens were fixed on a metallic support with a conductive adhesive carbon tape (Plano GmbH, Wetzlar, Germany), sputter-coated with a thin platinum layer (vacuum sputter coater SCD 050; Leica, Austria; thickness of the Pt layer: approx. 4 nm), and observed in an SEM microscope MAIA3 (Tescan, Brno, Czech Republic) using secondary electron imaging at accelerating voltage of 3 kV.

#### 2.4.2 Vibrational spectroscopy

Fourier-transform infrared spectra (FTIR) spectra were recorded on a Thermo Fisher Scientific Nicolet iS50 FTIR spectrometer (Nicolet CZ s.r.o.; Prague, Czech Republic) using 4 cm$^{-1}$ resolution in the 400–4000 cm$^{-1}$ region (with KBr beamsplitter and Happ-Genzel apodization) by means of ATR (diamond crystal) technique. Standard ATR correction was applied. Raman spectra were collected on a dispersive micro Raman system MonoVista CRS+ (Spectroscopy & Imaging GmbH; Warstein, Germany) interfaced to an Olympus microscope (50x objective) equipped with 785 nm excitation laser, grating 150 g/mm, spectrograph aperture 50 μm slit and laser power 10.5 mW at the sample. The wavelength and intensity calibration of the spectrometer was done by the software-controlled auto alignment procedure using mercury and Ne-Ar lamps. Total amount of 150–300 spectra for each sample with exposure time 5 s per spectrum were collected. All spectra were subsequently baseline corrected according to singular value decomposition method [34], averaged and normalized (max-min).

#### 2.4.3 Wide-angle X-ray scatttering

Wide-angle X-ray scattering (WAXS) patterns were obtained with a Panalytical MPD system (Panalytical; Almelo, the Netherlands) with vertical goniometer, CoKα radiation, variable divergence slits (mostly fixed irradiated length of 10 mm) and 1D Pixcel detector. The specimens (powders, films, bulk samples) were always put on the so-called non-diffracting Si substrates giving low background. The ranges of 4-70 deg 2θ (CoKα) were taken with the total measurement time 1 hr. For comparison with other literature, the final diffractograms were recalculated so that they corresponded to more common CuKα radiation (the wavelengths of CoKα and CuKα were taken as 1.79 and 1.54 Å, respectively). The crystallinity (weight fraction of crystalline phase) was calculated by means of Fityk software [35].

### 2.4.4 Dynamic Mechanical Thermal Analysis

The thermomechanical properties of the TPS systems were measured in torsion by dynamic mechanical thermal analysis (DMTA) on specimens of rectangular platelet shape (40 mm × 10 mm × 2 mm), using an ARES G2 (TA Instruments, New Castle, DE, USA) in oscillatory mode, at a deformation frequency of 1 Hz. The deformation amplitude ranged from 0.01 to 3% (regulated automatically by the auto-strain function, in response to sample resistance). The investigated temperature range was from −90 to 140 °C, while the heating rate was 3 °C/min. The temperature dependences of the storage shear modulus (G′), of the loss modulus (G″), and of the loss factor tan(δ) were recorded.

### 2.4.5 Rheology

Rheological properties of the TPS systems were measured in shear on a strain-controlled ARES G2 rheometer (TA Instruments, New Castle, DE, USA) using a parallel plate fixture with a diameter of 30 mm (plates with cross-hatched surface to prevent slipping). The thickness of the specimens was 2 mm. At first, the linear viscoelasticity region (LVER) was determined in view of the dependence of the storage modulus on the strain amplitude, which was measured at 120 °C at a frequency of 1 Hz. Next, the frequency sweep experiments were performed in a frequency range from 0.1 to 100 rad/s at a strain amplitude of 0.05 % (always well within the LVER), and at a constant temperature of 120 °C. To ensure a uniform temperature in the specimen, all samples were equilibrated for 2 min prior to the start of each type of experiment. The frequency sweep was performed twice for each TPS specimen.

### 2.4.6 Microindentation hardness testing

Micromechanical properties were measured with instrumented microindentation hardness tester (MCT tester; CSM, Switzerland). The microindentation hardness testing (MHI) experiments were carried out using a Vickers method: a diamond square pyramid (with an angle between non-adjacent faces 136°) was forced against the flat surface of a specimen. The flat smooth surfaces for the testing were prepared by cutting from the 2 mm thick plates with a rotary microtome RM 2255 (Leica, Vienna, Austria). The micromechanical properties were deduced from the loading force, which was measured as s function of penetration depth. From each sample, three independent cut surfaces were prepared. For each measured surface, at least 10 independent measurements/indentations were made and the final results were averaged. As we repeated the measurement for each sample three times to verify the reproducibility, the final averaged values of all micromechanical properties represent more than 90 measurements (3 cut surfaces per sample × at least 10 indentation per surface × 3 repetitions = more than 90). The parameters of MHI measurements were as follows: maximal loading force $F_{max}$ = 500 mN, dwell time (time of maximal load) 60 s, and linear loading and unloading rates 15,000 mN/min (i.e. ~2 s to achieve and release $F_{max}$). The evaluated micromechanical properties were: indentation modulus ($E_{IT}$) proportional to macroscopic elastic modulus, indentation hardness ($H_{IT}$) proportional to macroscopic yield stress, Martens hardness ($H_M$) also referred as universal hardness, indentation creep ($C_{IT}$) related to the macroscopic creep, and elastic part of the indentation work ($\eta_{IT}$) defined as ratio of elastic deformation to total deformation. The calculations of $E_{IT}$ and $H_{IT}$, $E_{IT}$ were based on the theory of Oliver and Pharr [36], while the values of $H_M$, $C_{IT}$ and $\eta_{IT}$ are independent on the O&P theory [37]. The exact definitions of above-listed micromechanical properties can be found in textbooks on micromechanical properties [38,39] and more detailed description of the MHI measurements are also in our recent studies [40–42].

### 2.4.7 In situ measurements during the melt mixing

During the melt-mixing of TPS in laboratory kneader (section 2.3.2), we recorded the values of torque (TQ, moment of force; in Nm) and real processing temperature ($T$; in °C) as a function of processing time ($t$; in seconds). To ensure the maximal reproducibility of the measurements, the kneading chamber was filled with the same amount of material (75 g) in the same time (2 min). Moreover, the final values of TQ and $T$ were evaluated from the final part of the TQ-$t$ and $T$-$t$ curves, after ca 6 min, when the curves reached a plateau indicating that the mixed system achieved a steady state.

## 3. Results and discussion

### 3.1 *Morphology and homogeneity*

#### 3.1.1. Light and electron microscopy

Figure 1 compares representative polarized-light micrographs (PLM) of thermoplastic starches prepared by two different protocols (one-step SC vs. two-step SC+MM) with or without acid hydrolysis (AH). In PLM micrographs, the bright areas indicate anisotropic material. In the case of TPS materials, the bright spots corresponded to non-fully-plasticized starch granules that kept their semicrystalline structure [2].

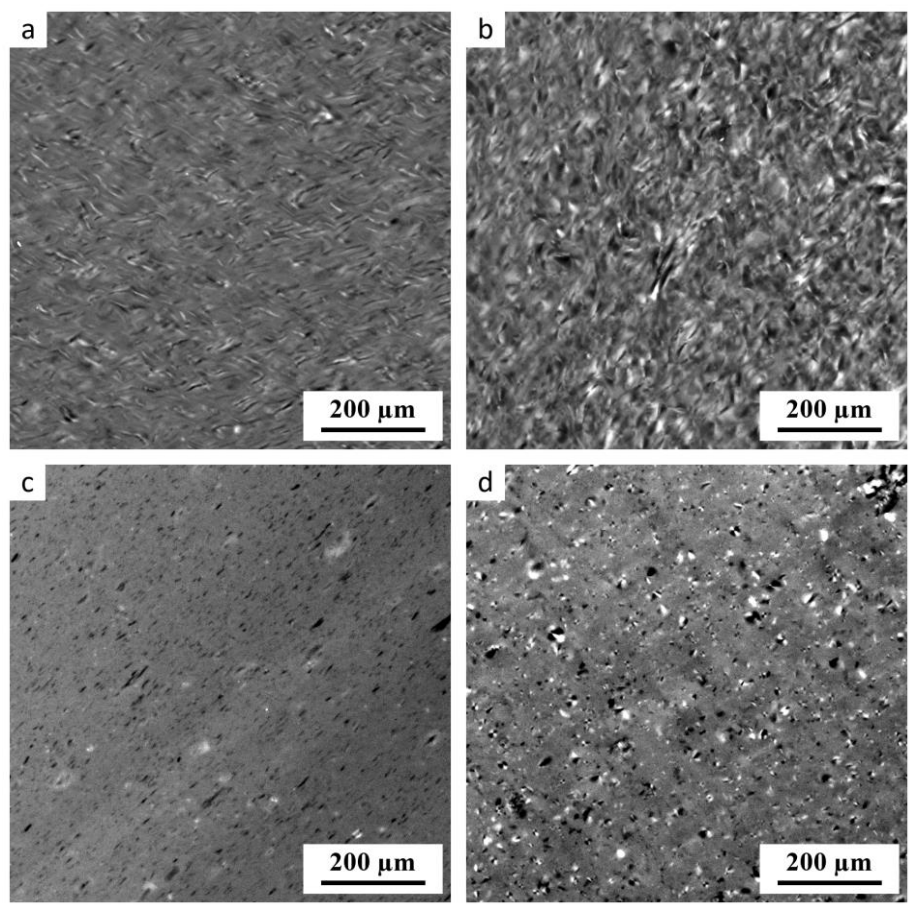

**Figure 1.** PLM micrographs showing four thermoplastic starch samples: (a) TPS-AH-0min after SC, (b) TPS-AH-60min after SC, (c) TPS-AH-0min after SC+MM, and (d) TPS-AH-60min after SC+MM. The samples without the AH treatment (Figs. a and c) contain less anisotropic inhomogeneities (bright spots) than the corresponding samples after AH (Figs. b and d). The complete list of the prepared TPS samples is given in Table 1.

The PLM results confirmed our previous findings [2,15,31] that the single-step SC does not yield highly homogeneous starch. The non-fully plasticized granules could be observed in TPS without AH treatment (Fig. 1a), and even more in the TPS after 60 min of AH (Fig. 1b). The most homogeneous starch was obtained the two step SC+MM preparation protocol without AH treatment (Fig. 1c). The two-step preparation after 60 min of AH resulted in less homogeneous material (Fig. 1d). This could be explained combining two facts known from the previous studies: (i) the amorphous regions of starch granules are more susceptible to AH than the crystalline regions [23,43] and (ii) if the amorphous matrix is more degraded and less viscous, the disintegration of starch granules is less complete due to the lower shear forces during SC+MM processing [2]. Interestingly, analogous trend is observed in immiscible polymer blends: if the viscosity of the matrix decreases, the disintegration of minor phase droplets during the melt mixing is less complete and the structure coarsens [44,45].

The morphology changes were monitored by an SEM microscopy as well. The SEM/SE micrographs of TPS fracture surfaces (Fig. A1 in Appendix A) confirmed the results of PLM (Fig. 1), but the differences among the samples could not be observed so clearly. The not-fully-plasticized starch granules exhibited higher contrast in polarized light (where they could be distinguished clearly as bright spots due to their anisotropic nature) than in SEM/SE micrographs (where they could be observed only in the form of unsharp, rounded asperities on fracture surfaces).

### 3.1.2. Vibrational spectroscopy

Figure 2 summarizes the vibrational spectroscopy results for all 12 studied samples, i.e. the three starch types (original powder, TPS after SC, and TPS after SC+MM) with four AH times (0, 15, 30, and 60 min). Both infrared spectroscopy (Fig. 2a) and Raman scattering (Fig. 2b) were in agreement that the dominant chemical change was the incorporation of glycerol into the starch structure during the SC, while the AH and MM did not alter the starch molecular structure significantly (see also Fig. A2 in Appendix A). The peak positions in TPS and their assignment to characteristic vibrations of starch and glycerol corresponded to those in the detailed study of Almeida et al. [46].

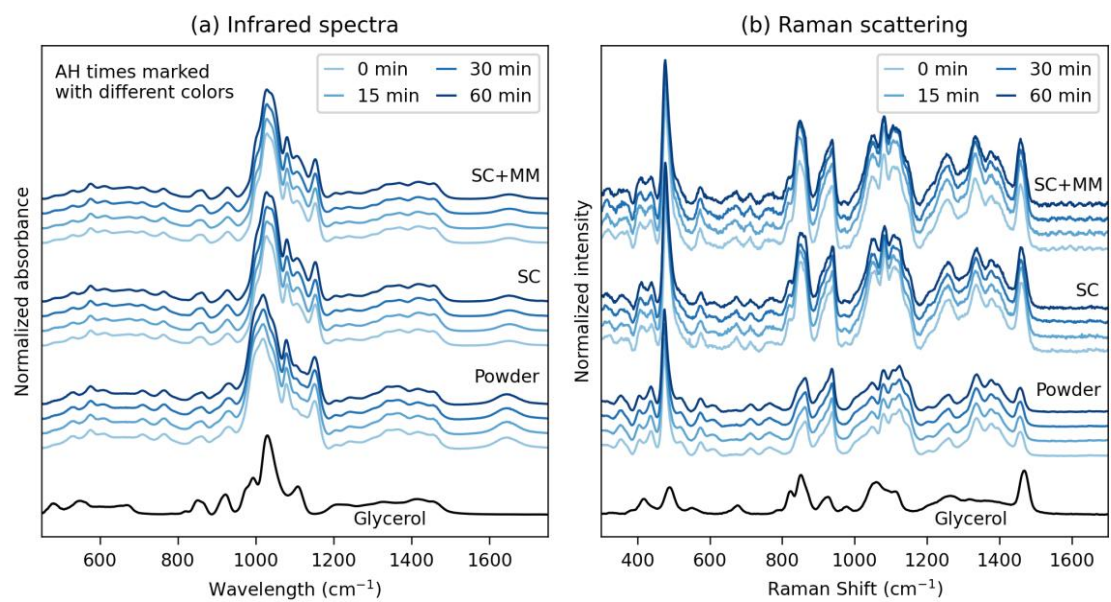

**Figure 2.** Vibrational spectroscopy results: (a) infrared spectra measured in ATR (attenuated total reflectance) mode and (b) Raman scattering measured at 785 nm excitation. Each of the two plots shows, from top to bottom, the following spectra: pure glycerol, original starch powders with various AH times, TPS after single-step SC preparation with various AH times, and TPS after two-step SC+MM preparation with various AH times.

There was a clear difference between the spectra of all original starch powders (glycerol-free materials) and the spectra of all TPS's (with glycerol added during the SC). The change of the spectra could be attributed to the fact that the glycerol molecules penetrated into the granules and interacted with starch molecules, forming new intra- and intermolecular hydrogen bonds [47]. Within each group of materials (powders, TPS after SC, and TPS after SC+MM), negligible variations were observed with the AH time. We conclude that the AH caused starch chain scissions changed both morphology (as discussed above) and properties (as discussed below), but the concentration of the newly formed end groups was below the detection limit of IR and Raman spectroscopy. This accorded with the literature [48], even if Chung et al. [49] have demonstrated that near-infrared (NIR) spectra are more sensitive in this case, enabling to monitor of the extent of AH. The differences between spectra of all TPS's after SC and after SC+MM were insignificant, although there were some local variations (we note that each spectrum Fig. 5 is a normalized average of >150 individual spectra, as described in the Experimental section).

### 3.1.3. Wide-angle X-ray scattering

Figure 3 shows the results of wide-angle X-ray scattering (WAXS) of all investigated samples as a function of increasing acid hydrolysis time. The native starch powders (Fig. 3, upper row) possessed the highest crystallinity. TPS after one-step SC protocol (Fig. 3, middle row) exhibited slightly lower crystallinity due to partial disintegration of native starch granules after gelatinization. TPS after two-step SC+MM protocol (Fig. 3, lower row) showed the lowest crystallinities due to (almost) complete destruction of the starch granules after the melt mixing. With increasing acid hydrolysis time, the crystallinity slightly increased (see all three rows of Fig. 3 from left to right). This accords with literature, which documents that the AH of starch preferentially targets the amorphous regions, enhancing crystallinity and double-helical content [23,25]. The crystallinity values for our investigated samples, ranging from 14 to 30 %, are quite in line with other available reports [29,50]. The increase in crystallinity after chain scissions in amorphous region is a general trend that has been observed also for other semicrystalline polymers, such as ultra-high molecular weight polyethylene (UHMWPE) after oxidative degradation [51,52].

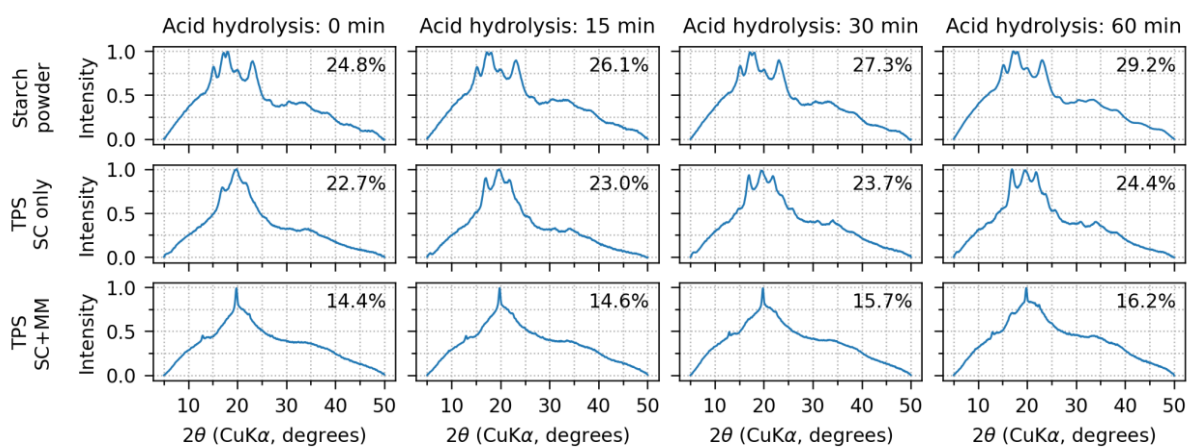

**Figure 3.** WAXS diffraction patterns and crystallinities of all investigated samples. The crystallinity values are printed in the upper right corner of each subplot. The rows show, from top to bottom: original starch powder, TPS after solution casting (SC), and TPS after solution casting and melt mixing (SC+MM). The columns show, from left to right, the studied materials after 0, 15, 30, and 60 min of acid hydrolysis.

All starch powder diffractograms (Fig. 3, upper row, all AH times) exhibited prominent diffraction peaks around $2\theta \approx 15°$, $17°$, $18°$, and $23°$ at CuK$\alpha$ wavelength. Moreover, several lower-intensity peaks could be observed at higher angles around $27°$, $31°$, $33°$, $38°$, and $41°$. All above-listed peaks are characteristic of the starch type-A crystallinity, which is typical of cereals [25,29,30,53]. Therefore, the results documented that the AH alone attacked mostly the amorphous regions and did not change the starch crystallinity type. Nevertheless, after SC the crystallinity started to change. Some diffractions decreased and/or disappeared due to partial destruction of the original amylopectin A-type crystallinity. Some new diffraction appeared, corresponding to newly formed amylose V-type crystals [2,53] and amylopectin B-type crystals formed due to A-to-B-type crystallinity transition [53,54]. After SC+MM, the crystallinity changed substantially. The diffractions of the original starch powder either disappeared completely or merged into broad peaks in regions $15°$–$25°$ and $30°$–$45°$. The only sharp diffractions at $13.5°$, $20°$, and $21°$ corresponded to V-type crystallinity [2,50]. The table summarizing all observed diffraction peaks together with the figure showing selected diffraction patterns with annotated diffractions are given in Appendix A (Table A1 and Figure A3).

## 3.2 *Mechanical and rheological properties*

The final TPS samples after SC+MM with various degrees of AH were characterized in detail by thermomechanical (Fig. 4), rheological (Fig. 5), and micromechanical measurements (Figs. 6 and 7). All three methods were in agreement that the properties of all samples were quite similar, although a slight increase in stiffness with increasing AH time was observed. The similar properties of the final TPS samples could be explained as follows: (i) the AH treatment caused chain scissions in the amorphous region, (ii) the chain scissions lead to softer and less viscous amorphous phase, (iii) the lower viscosity of the amorphous phase resulted in less complete disintegration of crystalline phase during solution casting and melt mixing as documented by PLM and WAXS results above, and (iv) the two contradictory effects – the softer amorphous phase with shorter molecules and the higher volume fraction of less disintegrated crystalline phase – tended to cancel out, although the impact of increased crystallinity prevailed moderately at the end. The increase in TPS stiffness after AH was observed in previous studies as well [30,55].

### 3.2.1. Dynamic mechanical thermal analysis

Figure 4 displays thermomechanical properties of the final TPS samples after SC+MM. The data were obtained from oscillatory shear rheometry (section 2.4.4) as temperature sweeps ($T$ from $-90$ to $140$ °C) at constant frequency (1 Hz). All four samples showed almost identical behavior. For storage modulus ($G'$; Fig. 4a) and loss modulus ($G''$; Fig. 4b) the differences are even hard to differentiate in the logarithmic scale plots. Nevertheless, the insets in Fig. 4a and 4b document that the acid hydrolysis tended to increase the values of $G'$ and $G''$ slightly in the region of laboratory temperature around 20 °C. This agreed quite well with the results of micromechanical measurements (Figs. 6 and 7 below).

The damping factors (tan δ; Fig. 4c) of all four samples exhibited three peaks around -50 °C, +30 °C, and +75 °C. These peaks corresponded to multiple glass transition temperatures ($T_g$) of TPS, as explained in detail elsewhere [2]. Briefly, the lowest-temperature peak is linked to the highly plasticized regions with high concentration of low-molecular-weight plasticizers (occasionally called *plasticizer-rich phase*), the intermediate peak is linked to less plasticized regions with higher concentration of starch molecules (occasionally called *starch-rich phase*), and the highest-temperature peak can appear if the starch-rich phase contains fragments of stiff semicrystalline structures. The exact shape of the DMTA curves depends on starch type, plasticization protocol, additives, humidity, aging etc., but we can conclude that our results were in reasonable agreement with previous studies [15,56,57]. More details concerning the interpretation of DMTA curves of TPS can be found in the abovementioned work of Rana et al. [2] and references therein. Complete results of DMTA results are in the Supplementary materials

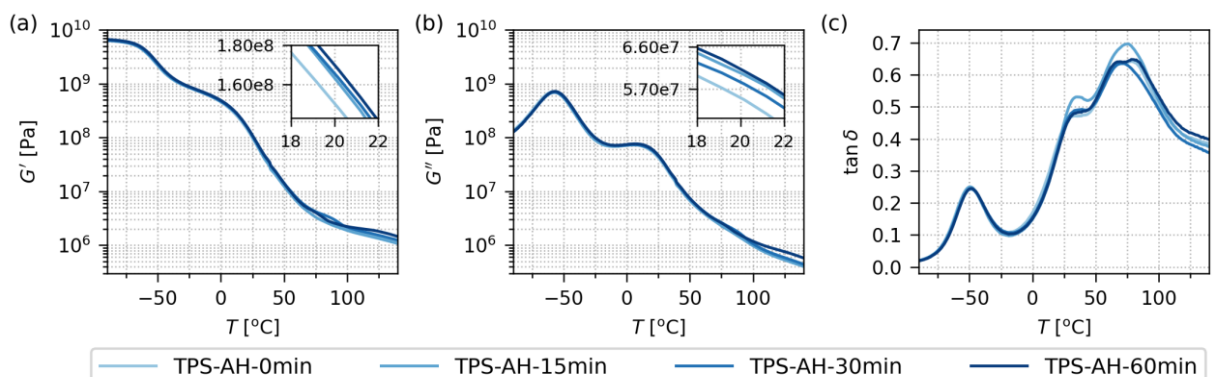

**Figure 4.** Thermomechanical properties of all investigated TPS samples after solution casting and melt-mixing, measured by DMTA in rectangular torsion mode in the temperature range from -90 to 140 °C: (a) storage modulus, $G'$, (b) loss modulus $G''$, and (c) damping factor, tan($\delta$). As the properties of all four samples were quite similar and hard to differentiate in logarithmic scale, the insets in Figs. (a) and (b) show the properties at laboratory temperature (20 °C) at higher detail. Note that the $G'$ and $G''$ are plotted with the same y-axis limits in order to facilitate the direct comparison of the two quantities.

### 3.2.2. Rheological properties

Figure 5 shows the rheological properties of the final TPS samples after SC+MM. The data were obtained from oscillatory shear rheometry (section 2.4.5) as frequency sweeps (angular frequency, $\omega$, ranging from 0.1 to 100 rad/s) at constant temperature (120 °C). The selected measurement temperature corresponded to the *nominal* TPS processing temperature, which was preset before the melt mixing (section 2.3.2). The extremely high molar mass of starch molecules caused that all samples exhibited a behavior typical of cross-linked polymers in the whole frequency range [17]: (i) $G' > G''$ in the whole range (solid and stiff network) and (ii) $|\eta^*|$ decreasing monotonously from "infinitely high" values (no zero-shear viscosity plateau typical linear polymers with common $M$).

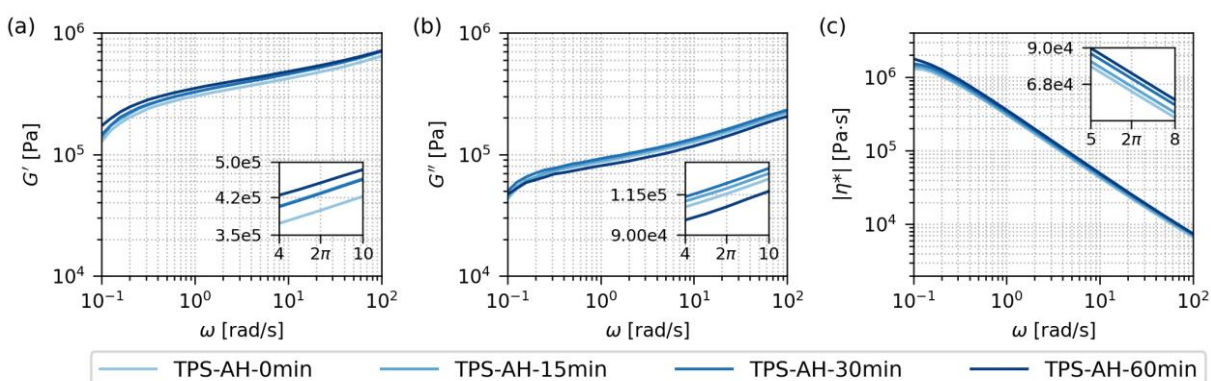

**Figure 5.** Rheological properties of all investigated TPS samples after solution casting and melt-mixing, measured by oscillatory shear rheometry at 120 °C: (a) storage modulus, $G'$, (b) loss modulus $G''$, and (c) absolute value of complex viscosity, $|\eta^*|$. As the properties of all four samples were quite similar and hard to differentiate in logarithmic scale, the insets show the properties at angular frequency $\omega = 2\pi$ at higher detail. The angular frequency $2\pi$ corresponds to frequency 1 Hz that was employed in DMTA experiments (Fig. 4; $f = 2\pi\omega$). The $G'$ and $G''$ are plotted with the same y-axis

limits in order to facilitate the direct comparison of the two quantities. <mark>Due to the extremely high molar mass of starch, the storage and loss modulus curves showed no intersection ($G' > G''$ a in the whole frequency range) and the $|\eta^*|$ exhibited a monotonous decrease (shear thinning without a plateau at low shear range).</mark>

In analogy with DMTA, the frequency sweeps in Fig. 5 confirmed the similar behavior of all TPS samples, regardless of the AH time. The insets in each plot of Fig. 5 show the minute differences between the samples at angular frequency $\omega = 2\pi$ rad/s, which corresponded to the frequency $f = 1$ Hz that was employed in DMTA (reminder: $\omega = 2\pi f$). The comparison of the insets in the temperature sweep plots (Fig. 4a–b) and frequency sweep plots (Fig. 5a–b) confirmed that the AH slightly increased the TPS resistance to elastic deformation ($G'$ increasing with AH time in the insets of Figs. 4a and 5a), while the TPS resistance to viscous flow was influenced much less ($G''$ exhibiting similar values and no clear trends in the insets of Figs. 4b and 5b). The absolute value of complex viscosity, $|\eta^*|$ showed a moderate increase with AH time (Fig. 5c). As the AH was expected to decrease the viscosity of TPS, this result is somewhat counter-intuitive. The simple explanation consists in that the $|\eta^*|$ is proportional to both $G'$ and $G''$ (reminder: $|\eta^*| = |G^*|/\omega = (G'^2+G''^2)^{1/2}/\omega$) and $G'$ increased with AH reproducibly, while $G''$ values did not change significantly. The more detailed explanation follows in section 3.3 that compares the TPS behavior in model conditions (an oscillatory flow in a rheometer) and in experimental conditions (a real flow during the melt mixing in a laboratory kneader).

### 3.2.3. Micromechanical properties

Figures 6 and 7 summarize the results of instrumented microindentation hardness testing (MHI) measurements for the four final TPS samples after SC+MM preparation. The micromechanical properties assessed from MHI can be divided into two groups [39,40]: (i) stiffness-related properties, such as indentation modulus, $E_{IT}$, indentation hardness, $H_{IT}$, and Martens hardness, $H_M$, and (ii) viscosity-related properties, such as indentation creep, $C_{IT}$, and elastic part of the indentation work, $\eta_{IT}$. The main results connected with the stiffness-related properties are shown in Fig. 6, the main results linked with viscosity-related micromechanical properties are shown in Fig. 7, and the exact definitions of all above-listed micromechanical properties can be found in Appendix B (Fig. A4).

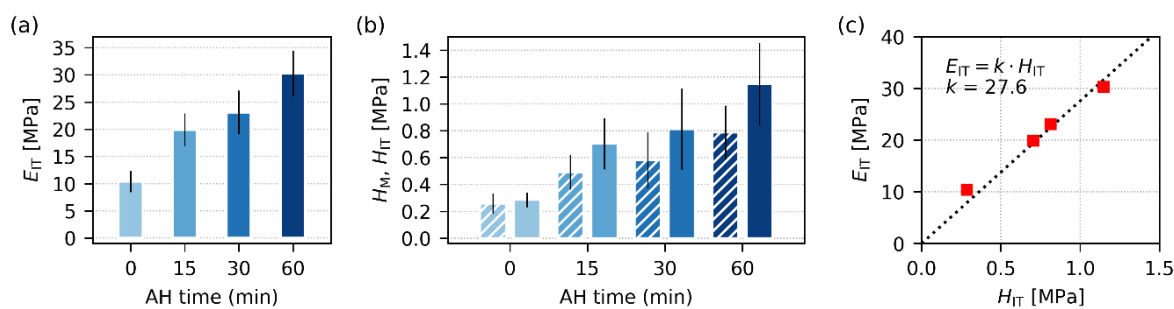

**Figure 6.** Stiffness-related properties of all investigated TPS samples after SC+MM, obtained from microindentation hardness testing: (a) indentation modulus, $E_{IT}$, (b) Martens hardness ($H_M$; striped columns) and indentation hardness ($H_{IT}$, standard filled columns), and (c) the theoretically predicted linear correlation between $E_{IT}$ and $H_{IT}$. For each sample, we performed at least 90 indentations and averaged the results; the error bars represent estimated standard deviations.

The MHI measurements of stiffness-related properties (Fig. 6) confirmed that the AH enhanced the overall stiffness of TPS. The clear monotonous increase with AH time was observed for $E_{IT}$ (Fig. 6a) as well as for $H_{IT}$ and $H_M$ (Fig. 6b). This accorded with both DMTA (Fig. 4a) and rheological measurements (Fig. 5a), in which another stiffness-related property, $G'$, increased with AH time as well. The trends in MHI (Fig. 6) look clearer and more reproducible than in the case of DMTA and rheology (Figs. 4 and 5). This resulted from one important advantage of micromechanical measurements: Due to the small size of indentations (typically around 100 μm), we can perform tens or even hundreds of measurements within one specimen. In our case, the temperature sweeps (Fig. 4) represent average of 2 independent measurements, the frequency sweeps (Fig. 5) are average of 9 experiments, and micromechanical properties (Figs. 6 and 7) come from more than 90 independent measurements as explained in section 2.4.6. Another confirmation of the reliability and reproducibility of the MHI measurements is displayed in Fig. 6c. Both theoretical considerations [58,59] and previous experimental results [39,60] suggest that many polymer systems exhibit approximate linear correlations between all stiffness-related properties. In our case, the theoretically predicted linear correlation between $E_{IT}$ and $H_{IT}$ was almost perfect, as documented in Fig. 6c.

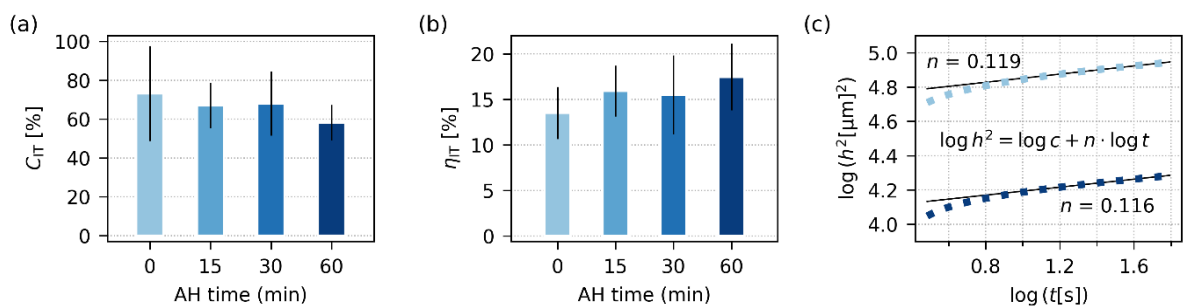

**Figure 7.** Viscosity-related properties of all investigated samples TPS samples after SC+MM, obtained from microindentation hardness testing: (a) indentation creep, $C_{IT}$, (b) elastic part of the indentation work $\eta_{IT}$, and (c) the theoretically predicted linear correlation between squared indenter penetration depth ($h^2$) time ($t$) in log-log scale. In Fig. (c) the light blue squares represent TPS-AH-0min sample, the dark blue squares represent TPS-AH-60min sample, and the thin black lines are linear regression curves. For each sample, we performed at least 90 indentations and averaged the results; the error bars represent estimated standard deviations.

The viscosity-related properties (Fig. 7) were affected by the AH less than the stiffness-related properties. This corresponded to DMTA and rheological measurements (Figs. 4 and 5), in which the viscosity-related loss moduli, $G''$, were influenced less than the stiffness-related storage moduli, $G'$. The values of $C_{IT}$ and $\eta_{IT}$ just slightly decreased and increased with AH time, respectively. These shifts can be attributed to the fact that AH increases the crystallinity and the TPS with higher content of crystalline phase are stiffer (higher $E_{IT}$ and $H_{IT}$), more resistant to creep (lower $C_{IT}$), and more elastic (higher $\eta_{IT}$). Analogous behavior was observed for other semicrystalline polymers such as polypropylene [61] or poly(lactic acid) [37]. Figure 7c re-confirms the correctness of microindentation measurements, as the creep exhibits the typical power-law behavior, i.e. the typical linear increase in log($h^2$)-log($t$) scale, as explained elsewhere [62]. Moreover, the values of creep constants, $n$, which were determined from fitting the power law to experimental data, are in good agreement with literature [62,63].

### 3.3 Processing of acid hydrolyzed starches

Figure 8 shows the results of *in situ* measurements during the melt-mixing of TPS. The melt-mixing was the final preparation step of our thermoplastic starches, which resulted in the most homogeneous materials, as illustrated in Fig. 1. The main objective of this work was to verify if the acid hydrolysis can decrease the processing temperature during the final melt-mixing step. We note that the *nominal processing temperature*, which is preset in the laboratory kneader before the mixing (120 °C, as described in section 2.3.2) somewhat increases after filling the mixing chamber with TPS due to the internal friction and energy dissipation [2]. As a result, during the melt-mixing we can measure the energy needed to make TPS flow in the chamber (TQ, torque, moment of force; Figs. 8a–b) and the increased *real processing temperature* ($T$; Figs. 8c–d) as described in section 2.4.7.

Figure 8 confirms that the AH could decrease both torque and processing temperature. The observed decrease was quite moderate for both quantities (Figs. 8a and 8c), but even a few degrees can save considerable amount of energy in large-scale applications [64,65] or improve the persistence of temperature-sensitive admixtures, such as antibiotics [31,33]. The TQ-*t* curves (Fig. 8b) and *T*-*t* curves (Fig. 8d) exhibited a steep change when TPS was added to the chamber during the first 2 min: TQ increases from zero (as we need enough energy to mix the cold and stiff material), $T$ decreased from the nominal 120 °C (as the source TPS, which was kept at ambient temperature, cooled the mixing chamber). After ca 6 min of mixing, the TPS systems achieved the steady state (a plateau on TQ-*t*, and *T*-*t* curves) where TQ decreased (the heated and homogenized material required less energy to flow) and $T$ increased (a dissipation of energy due to internal friction of the flowing viscous material in the filled chamber). The average values of TQ and $T$ in Figs. 8a and 8c were taken from the plateau regions of TQ-*t* and *T*-*t* curves in Figs. 8b and 8d, respectively.

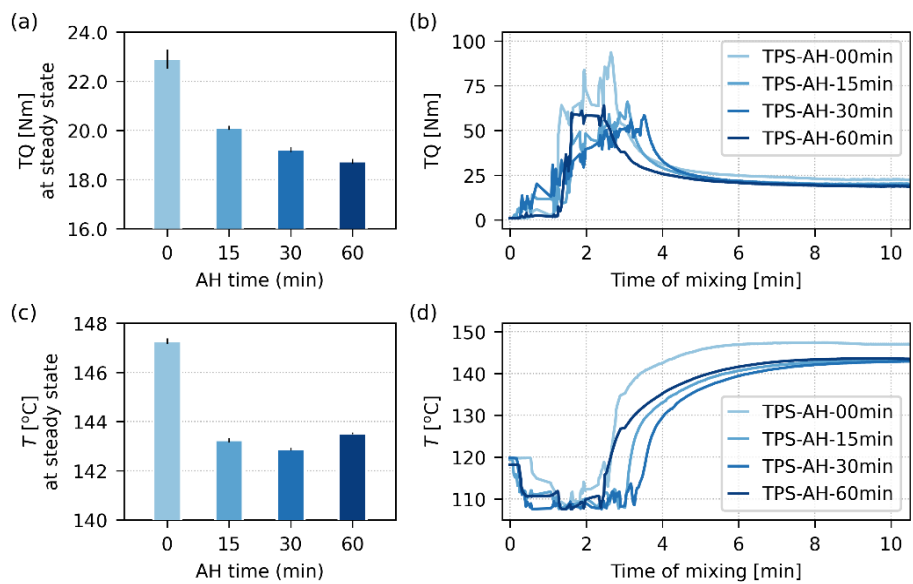

**Figure 8.** Processing parameters during the melt mixing of TPS-AH systems, which were measured *in situ* during the sample preparation: (a, b) torque (TQ) and (c, d) processing temperatures ($T$). The bar plots on the right (a, c) show the final average values of TQ and $T$ at steady state, i.e. after ca 6 min of melt mixing. The line plots on the left (b, d) show TQ and T a function of time during the whole melt mixing process. The mixing chamber was pre-heated to the *nominal* processing temperature of 120 °C, while the *real* processing temperature ($T$) was measured *in situ* as shown in (d); the real temperature increase was connected with the internal friction energy dissipation.

If we compare the rheological behavior of TPS at model conditions during oscillatory shear rheometry (Fig. 5) with the TPS behavior during the real melt-mixing in the kneader (Fig. 8), we can see a discrepancy that illustrates the limitations of simplified rheological measurements. In the rheometer, all TPS samples behaved as a *solid material* (Figs. 5a–b: $G' > G''$ in the whole frequency range) and the AH time *increased* their viscosity (Fig. 5c). In the kneader, all TPS samples behaved a *viscous liquid* (as they could flow in the chamber) and the AH decreased their viscosity (as evidenced by the decrease of TQ and *T* for the AH-treated samples with respect to non-modified TPS). The explanation of this seeming contradiction is as follows: (i) The oscillatory shear rheometer characterizes TPS at 120 °C in the linear viscoelasticity range (section 2.4.5). (ii) At the constant temperature of 120 °C and moderate oscillatory flow, the material keeps its structure and behaves as a solid. (iii) The kneader forces TPS to continuous movement, which increases the temperature due energy dissipation and disrupts many molecular entanglements together with remaining semicrystalline agglomerates. (iv) Under these conditions, the material starts to flow and behaves as a viscous liquid. (v) Once the material is under continuous flow, the shorter molecules present due to the AH start to act as lubricants that decrease the overall viscosity and the final processing temperature. We conclude that in our specific case of TPS with its extremely long molecules and semicrystalline aggregates, the simple oscillatory shear rheometry could gave us the basic characterization of rheological behavior (as summarized in Fig. 5), but it could not yield a realistic prediction of the rheological behavior during the real melt-mixing experiment (as documented in Fig. 8).

## 4. Conclusions

The main objective of this work was to verify if the acid hydrolysis (AH) can decrease the processing temperature of thermoplastic starch (TPS) during the melt mixing. We subjected native wheat starch to 0, 15, 30, and 60 min of AH and converted the modified materials to highly-homogeneous TPS by means a two-step preparation protocol, which comprised solution casting (SC) followed by melt-mixing (MM). All prepared TPS's were thoroughly characterized by multiple methods including polarized light microscopy (PLM), vibrational spectroscopy (infrared and Raman), wide-angle X-ray scattering (WAXS), dynamic mechanic thermal analysis (DMTA), oscillatory shear rheometry, instrumented microindentation hardness testing (MHI), and *in situ* measurements of torque and temperature during the final MM step. The key results can be summarized below:

- The acid hydrolysis preferentially targeted the amorphous regions, which decreased the average molar mass and viscosity of the amorphous matrix. During the SC and MM, the lower-viscosity matrix in the AH-treated starches resulted in lower disruption of semicrystalline regions, slightly coarser morphology (as observed by PLM) and higher crystallinity (as evaluated from WAXS).
- The fact that TPS after AH exhibited less viscous softer amorphous phase and higher fraction of stiffer crystalline phase lead to similar thermomechanical, rheological, and micromechanical properties of all prepared systems (as evidenced by DMTA, rheometry, and MHI) because the two phenomena tended to cancel out.
- During the melt mixing, when the material was fully molten and forced to flow, the shorter molecules in all AH-treated TPS started to act as a lubricant and decreased both torque (TQ) and processing temperature (*T*), as proved by *in situ* measurements in the kneading chamber during the MM. Even if the changes of TQ and *T* were quite moderate, we have demonstrated that the AH is a feasible approach to save energy during the TPS processing and/or to mitigate the negative effect of the MM on possible temperature-sensitive admixtures.

**Author Contributions:** Conceptualization, M.S.; validation, S.K., V.G. and M.S.; formal analysis, S.K., V.G., I.S. and M.S.; investigation, S.K., V.G., B.S., I.S., R.S. and Z.S.; resources, B.S., R.K., Z.S. and M.S.; data curation, S.K., V.G. and M.S.; writing—original draft preparation, S.K., V.G., B.S., I.S.,

R.K. and M.S.; writing—review and editing, M.S.; visualization, S.K., V.G. I.S. and M.S.; supervision, M.S.; project administration, M.S.; funding acquisition, M.S. All authors have read and agreed to the published version of the manuscript.

**Funding:** This research was funded by Technology Agency of the Czech Republic, program NCK2, grant number TN02000020. I.S. acknowledges the assistance provided by the Advanced Multiscale Materials for Key Enabling Technologies project, supported by the Ministry of Education, Youth, and Sports of the Czech Republic. Project No. CZ.02.01.01/00/22_008/0004558, Co-funded by the European Union.

**Institutional Review Board Statement:** Not applicable.

**Data Availability Statement:** The data are available at request to the corresponding author.

**Acknowledgments:** M.S., S.F. and V.G. would like to thank Pavel Nemecek for his assistance with LM and SEM measurements and Anastasiya Rakitina for her assistance with melt-mixing of thermoplastic starch. I.S. acknowledges CUCAM Centre of Excellence (OP VVV "Excellent Research Teams" project No. CZ.02.1.01/0.0/0.0/15_003/0000417) for the purchase of MonoVista Raman spectrophotometer.

**Conflicts of Interest:** The authors declare no conflicts of interest.

## Abbreviations

The following abbreviations are used in this manuscript:

| | |
|---|---|
| TPS | Thermoplastic starch |
| SC | Solution-casting |
| MM | Melt-mixing |
| SC+MM | Solution-casting followed by melt-mixing |
| AH | Acid hydrolysis |
| LM | Light microscopy |
| PLM | Polarized light microscopy |
| SEM | Scanning electron microscopy |
| FTIR | Fourier-transform infrared spectra |
| WAXS | Wide-angle X-ray scattering |
| DMTA | dynamic mechanical thermal analysis |
| LVER | linear viscoelasticity region |
| MHI | microindentation hardness testing |

## Appendix A: Additional TPS characterization results

Figure A1 shows SEM micrographs of fractures surfaces of thermoplastic starches prepared by two different protocols (SC vs. SC+MM) with or without 60 min of acid hydrolysis. Briefly, the SEM micrographs confirmed the conclusions from PLM micrographs (Fig. 1), but the morphology changes were not so clearly visible. This was caused by the lower contrast on the SEM micrographs (topographic contrast on fracture surfaces) in comparison with the higher contrast in PLM micrographs (polarized light shows anisotropic regions in the non-fully plasticized starch granules as bright spots).

Figure A2 displays detailed comparison of four Raman spectra: pure glycerol, wheat starch powder, TPS after single step SC preparation and TPS after two-step SC+MM preparation. The comparison documents that that the difference between vibration spectra of SC and MM samples was negligible. A detailed Raman analysis of starch powder with the tentative band assignments was done by Almeida et al. [46]. Our results are in a good agreement with this work. Moreover, they indicate that spectral shifts of SC and SC+MM

Table A1 summarizes all diffraction peaks observed in the 12 measured samples (i.e. powders, SC samples, and SC+MM samples after 0, 15, 30, and 60 min of AH). Figure A3 shows selected diffraction patterns (i.e. all powders, SC samples, and SC+MM samples after 0 and 60 min of AH) with annotated diffraction peaks. The assignment of the diffraction peaks together with the relevant references and brief description is given in Table A1.

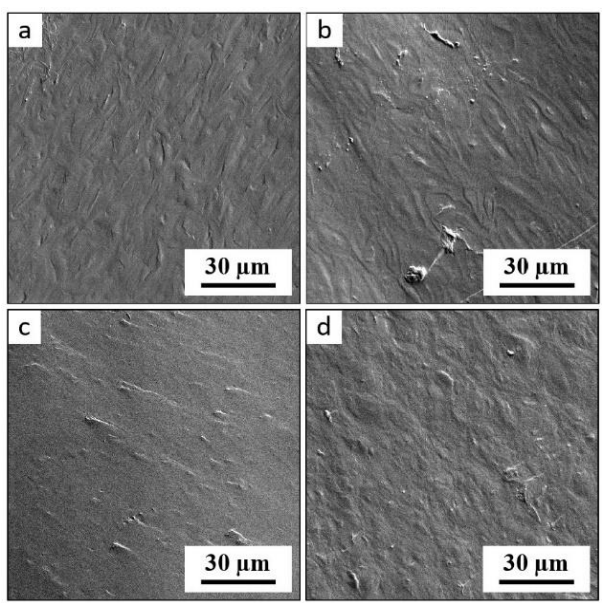

**Figure A1.** SEM micrographs showing fracture surfaces of TPS samples: (a) TPS-AH-0min after SC, (b) TPS-AH-60min after SC, (c) TPS-AH-0min after SC+MM, and (d) TPS-AH-60min after SC+MM. The list of the prepared TPS samples is given in Table 1; SC = solution casting; SC+MM = solution casting followed by melt mixing. The samples were observed in the secondary electron mode at the accelerating voltage of 3 keV.

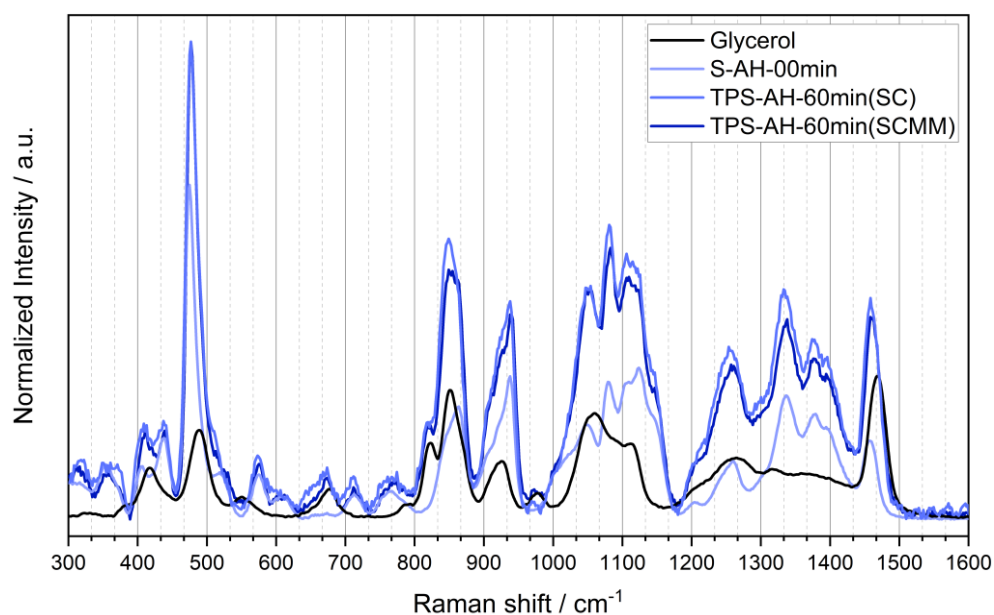

**Figure A2.** Raman spectra of four studied samples: pure glycerol, wheat starch powder, TPS after single step SC and 60 min of AH and TPS after two-step SC+MM and 60 min of AH.

**Table A1.** Summary of WAXS diffraction peaks.

| WAXS diffractions | | Observation of the diffraction peaks | | | Starch structure types | |
|---|---|---|---|---|---|---|
| 2θ(°) | Intensity [1] | S(powder) | TPS(SC) | TPS(SC+MM) | Structure [2] | References |
| 5.3 | w | no | yes | no | B | [2] |
| 13.5 | m | no | no | yes | V | [50,53] |
| 15.1 | s | yes | no | no | A/B | [30] |
| 17.0 | s | yes | yes | no | A/B | [2,30] |
| 18.1 | s | yes | no | no | A | [30] |
| 19.7 | s | no | no | yes | V | [50,53] |
| 20.0 | s | yes | yes | no | A/B | [2,30] |
| 20.8 | w | no | no | yes | V | [50,53] |
| 21.8 | m | no | yes | no | V | [2] |
| 23.2 | w | yes | no | no | A | [30] |
| 26.6 | m | yes | yes | no | A/B | [30] |
| 30.6 | w | yes | yes | no | r | [29,66,67] |
| 33.4 | w | yes | yes | yes | r | [29,66,67] |
| 38.4 | w | yes | yes | yes | r | [29,66] |
| 43.1 | w | yes | no | no | r | [29,66] |

[1] WAXs diffraction intensity notation: w = weak; m = medium; s = strong

[2] Starch structure type: A = residual A-type amylopectin crystallinity from the original starch powder; B = newly formed B-type amylopectin crystallinity due to intensive hydration that occurs during the SC; V = newly formed $V_H$-type amylose crystallinity according to [53] , r = residual crystallinity from the original starch powder, lower intensity peaks at higher angles.

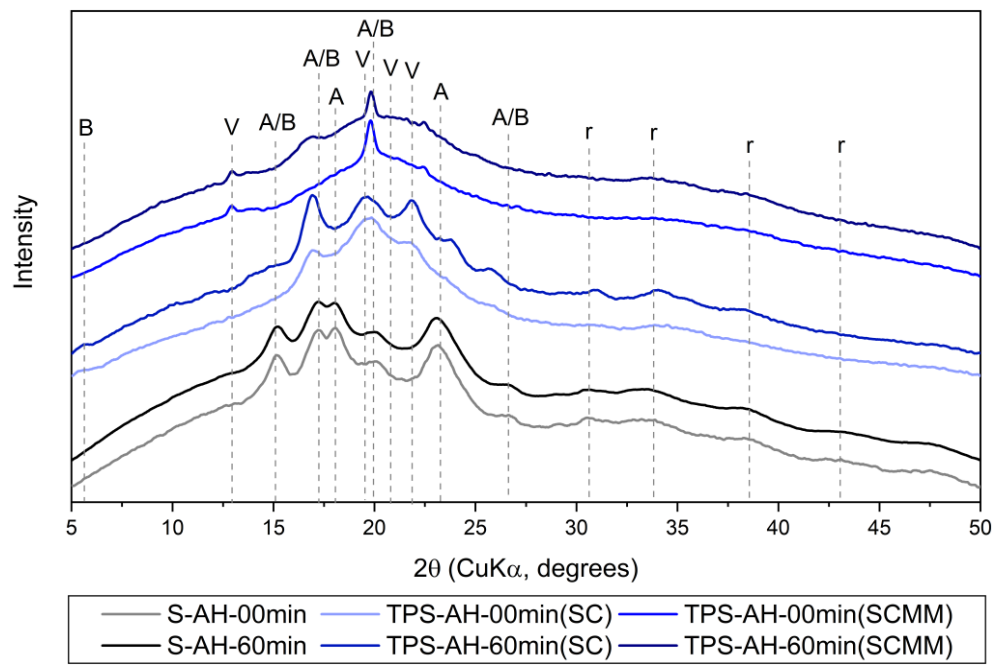

**Figure A3.** Selected WAXS diffraction patterns with annotated diffraction peaks. The diffractions are marked according to Table A1 and references therein.

## Appendix B. More details about TPS micromechanical properties

Figure A4 summarizes the principle of microindentation measurement, shows raw experimental data in the form of representative *F-h* curves, and gives exact definition of all micromechanical properties ($E_{IT}$, $H_{IT}$, $H_M$, $C_{IT}$, and $\eta_{IT}$) employed in this study. Complete results of micromechanical measurements are in the Supplementary materials.

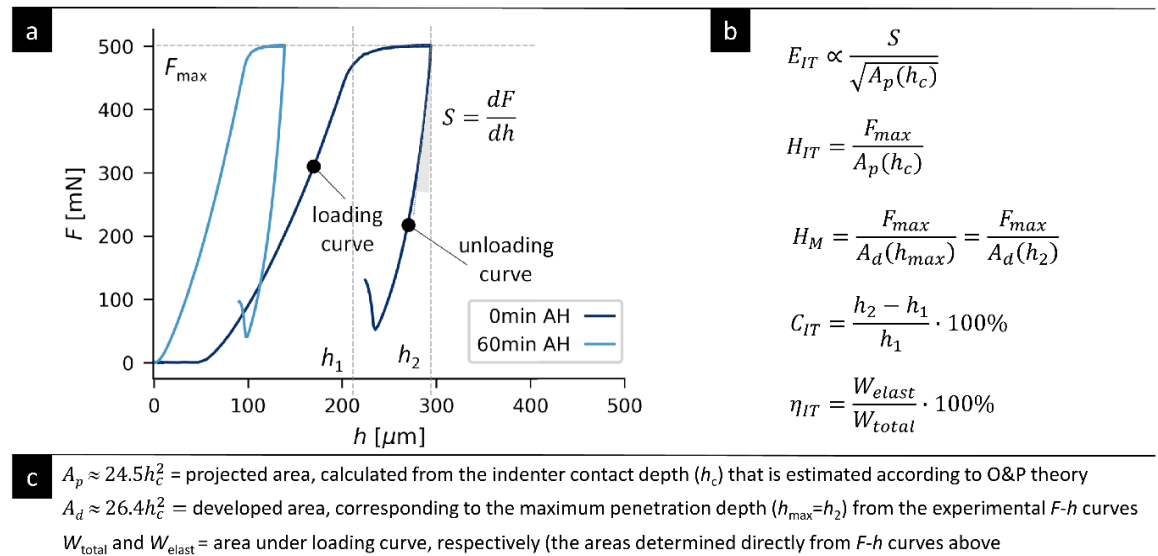

$A_p \approx 24.5 h_c^2$ = projected area, calculated from the indenter contact depth ($h_c$) that is estimated according to O&P theory

$A_d \approx 26.4 h_c^2$ = developed area, corresponding to the maximum penetration depth ($h_{max}=h_2$) from the experimental *F-h* curves

$W_{total}$ and $W_{elast}$ = area under loading curve, respectively (the areas determined directly from *F-h* curves above

**Figure A4.** Representative real *F-h* curves from microindentation hardness testing for TPS-AH-0min and TPS-AH-60min samples and brief description of principle of MHI measurements. (a) micromechanical properties were deduced from the *F-h* curves (*F* = indenter loading force, *h* = indenter penetration depth) by means of the formulas from (b) and the relationships from (c) containing experimental parameters, such as maximum loading force ($F_{max}$), slope at the beginning of the unloading curve (*S*), penetration depths at the beginning and end of the maximal load ($h_1$ and $h_2$), and areas under the loading and unloading curve ($W_{elast}$ and $W_{total}$). The additional parameter, contact depth ($h_c$), was calculated in terms of the Oliver & Pharr theory and employed in the calculation of $E_{IT}$ and $H_{IT}$.